\documentclass{article}
\usepackage{graphicx}
\usepackage{amsmath}
\usepackage{amsfonts}
\usepackage{amssymb}
\usepackage{indentfirst}
\newtheorem{theorem}{Theorem}

\newtheorem{lemma}[theorem]{Lemma}

\newtheorem{proposition}[theorem]{Proposition}

\begin{document}

\title{Noisy Grover's search algorithm\footnote{ICQI 2001 - International Conference on Quantum Information, Rochester, June 10-13, 2001 (Contribution given by D. E).
}}

\author{Demosthenes Ellinas and Christos Konstadakis
}
\maketitle
\begin{center}
Department of Sciences, Section of Mathematics \\
Technical University of Crete
\\

GR-73 100 Chania, Crete, Greece \\
ellinas@science.tuc.gr, xr@creteisland.gr
\end{center}

\begin{abstract}

External environment influences on Grover's search algorithm modeled
by quantum noise are investigated. The algorithm is shown to be robust under
that external dissipation. Explicitly we prove that the resulting search
positive maps acting on unsorted N-dimensional database made of projective
density matrices depend on x the strength of the environment, and that there
are infinitely many x values for which search is successful after O($\sqrt{N}$
) queries. These algorithms are quantum entropy increasing.
\\

\noindent *1999 Optical Society of America


\end{abstract}
\newpage

\section{Introduction
}

Let us consider Grover's algorithm with searching matrix U$_{G}=-UI_{s}%
U^{\dagger}I_{w}$, used to investigate for a single item $|w\rangle$ among N
orthonormal others, that span the complex Hilbert space of an unsorted quantum
database D =$\left\{  |i\rangle\right\}  _{i=1}^{N}$ , where U is a general
U(2) unitary matrix, and $I_{s}=1-2|s\rangle\langle s|$ , $I_{w}%
=1-2|w\rangle\langle w|$ \ are reflection operators wrt the vectors
\{$\ |s\rangle,|w\rangle$\}, while $|s\rangle=\frac{1}{\sqrt{N}}\underset
{i=1}{\overset{N}{\sum}}|i\rangle$ is the uniform superposition state of the
database [1-4]. The remarkable fact is that while classical search requires on
the average N trials for finding the target item w, the quantum algorithm is
quadratically faster since it determines $|w\rangle$ after only O($\sqrt{N}$ )
queries. Our aim here is to reconsider the algorithm and to extent the concept
of database searching to more realistic conditions where we take into account
the presence of an external environment in the form of quantum noise. Indeed
external influences on quantum processes and their induced decoherence and
dissipation are unavoidable, so their quantitative study is of fundamental
importance for the field of quantum information [5]. Such a study will be
carried out here utilizing the U(2) symmetry of Grover's algorithm[3, 4] by
first selecting $U=U_{\frac{\pi}{4}}$ to be $\frac{\pi}{4}$ -rotation which is
corrupted by external influences modeled as an interaction with a two
dimensional quantum environment of strength $\chi\geq0.$

\section{Construction of noisy search algorithm
}

Let the environment be described by a quantum system with 2D- Hilbert space
that perturbates the Hamiltonian of the $\frac{\pi}{4}$ -rotation by the
following interaction: $H_{R}=H_{\frac{\pi}{4}}+H_{env}=\frac{\pi}{4}\sigma
_{y}\bigotimes1+\frac{\chi}{2}(1-\sigma_{z})\bigotimes\sigma_{y}$ [6]. From
this Hamiltonian we obtain the evolution operator $U_{R}=$exp($iH_{R}$) ,
where x is the strength of interaction with the environment. This operator
signifies a noisy $\frac{\pi}{4}$ -rotation, where phase damping has been
incorporated into it. The net effect of noise on the rotation operator is to
replace it by the completely positive trace preserving map(CP) $\rho
\rightarrow\varepsilon^{^{\prime}}(\rho)=Tr_{env}(U_{R}$ $\rho\bigotimes
|0\rangle\langle0|U_{R}^{\dagger})=\underset{i=0}{\overset{1}{\sum}}R_{i}\rho
R_{i\text{ \ }}^{\dagger}$, where $R_{i}=Tr_{env}(U_{R}|i\rangle
\langle0|)=\langle i|U_{R}|0\rangle$ , are the so called Kraus operators of
the CP map satisfying the completeness relation \ $\underset{i=0}{\overset
{1}{\sum}}R_{i}^{\dagger}R_{i}=1.$ \ [7].

\begin{proposition}  The noisy $\frac{\pi}{4}$-rotation map $\rho\rightarrow
\varepsilon^{^{\prime}}(\rho)=\underset{i=0}{\overset{1}{\sum}}R_{i}\rho
R_{i}^{\dagger}$ , is determined by the operators

$R_{0}=(\cos\mu(\chi)1+\frac{i\chi}{2}\delta(\chi)\sigma_{y})e^{-\frac{ix}%
{2}\sigma_{y}}$ , $R_{1}=\frac{\pi}{4}\delta(\chi)e^{\frac{-i\chi}{2}%
\sigma_{y}}$ , \\ where 

$\delta\left(  \chi\right)  =\frac{\sin\mu\left( 
\chi\right)
 }{\mu\left(  \chi\right)  }$ , $\mu\left(  \chi\right) 
=\sqrt{\frac{\chi^{2}%
 }{4}+\frac{\pi^{2}}{16}}$, $\chi\geq0.$

\end{proposition}

\begin{lemma} Self-composition of unitary CP map gives unitary CP map.

\end{lemma}

At this point we perform an optimal unitary preconditioning on $\ \varepsilon
^{^{\prime}}\left(  \rho\right)  .$ We recall that the nearest wrt Euclidean
metric unitary matrix to a given square one, is the one involved in the so
called polar decomposition, namely in the decomposition given by the product
of a hermitian times a unitary matrix [8]. Then due to Lemma 2 we replace the
generators of the CP map $\varepsilon^{^{\prime}}\left(  \rho\right)  $ \ by
their respective nearest unitary ones and in this way we obtain the following
unitary CP map $\varepsilon\left(  \rho\right)  $ .

\begin{proposition} The unitary CP \ $\varepsilon$ is generated by the
unitaries
 \{$\frac{1}{\sqrt{2}}V_{0},\frac{1}{\sqrt{2}}V_{1}\}$ with
$V_{0}=e^{i\left(
 \psi\left(  \chi\right)  -\frac{\chi}{2}\right) 
\sigma_{y}}$ and
 $V_{1}=e^{-\frac{i\chi}{2}\sigma_{y}}$, where 
\begin{center}
\ $\left[ 
\cos^{2}\mu\left(
 \chi\right)  +\frac{\chi^{2}}{4}\delta^{2}\left( 
\chi\right)  \right]
$ $\cos^{2}\psi\left(  \chi\right)$ $=\cos^{2}\mu\left( 
\chi\right)  $.
 
\end{center}
\end{proposition}

We now introduce the following extension of Grover's algorithm. Let the
database $\Pi=\left\{  |i\rangle\langle i|\right\}  _{i=1}^{N}$ $=\{\rho
_{i}\}_{i=1}^{N}$, consisting from a collection of \ N pure density matrices
obtained by the state vector database D . For unitaries \ $X,Y$ let the adjoin
map $AdX:\Pi\rightarrow\Pi:\rho\rightarrow AdX\left(  \rho\right)  =X\rho
X^{\dagger}$ , with the property \ $AdXY\left(  \rho\right)  =AdX\left(
AdY\left(  \rho\right)  \right)  $. Then \ $AdU_{G}\left(  \rho_{s}\right)
=AdUAdI_{s}AdU^{\dagger}AdI_{w}=U_{G}\rho U_{G}^{\dagger}$ , is the
implementation of Grover's search map in the database $\Pi$ .

Let e. g $U=U_{\frac{\pi}{4}}$ , then the effect of the environment will
amount to replace the $\frac{\pi}{4}$-rotation by the unitary CP map
\ $\varepsilon$\ as was shown before. This in turn will cause the embedding of
$AdU_{G}$ \ into the unitary CP map $t:\Pi\rightarrow hull\left(  \Pi\right)
$ , that maps pure density matrices of the database $\Pi$ , to mixtures of
states that form the convex hull of elements of $\Pi$ . Explicitly we obtain
\ $t=\frac{1}{\sqrt{2}}AdV_{0}I_{s}V_{0}^{\dagger}I_{w}+\frac{1}{\sqrt{2}%
}AdV_{1}I_{s}V_{1}^{\dagger}I_{w}.$

\section{Grover's algorithm is robust under quantum noise
}

We now proceed employing the unitary CP search map $t^{m}\left(  \rho
_{s}\right)  =t^{m}\left(  |s\rangle\langle s|\right)  $, acting m times on
the initial pure density matrix $\rho_{s}$ . To quantify the complexity of
searching we note that the Bloch vector associated to the density matrices of
the database gives a more clear picture of searching. Indeed we can show that
for general noise parameter x the search map \ $t$ induces an exponential, wrt
the number of queries, damping in the norm of Bloch vectors. Therefore to
evaluate the efficiency of the algorithm we need two figures of merit, the
radial fidelity giving the projection between the Bloch vectors of target and
final density operators $\ f=\langle$ $t^{m}\left(  \rho_{s}\right)  ,\rho
_{w}\rangle=\frac{1}{2}Tr\left(  t^{m}\left(  \rho_{s}\right)  \rho
_{w}\right)  $ , and the cosine of angular fidelity between the same vectors
$\cos\gamma=\frac{\langle t^{m}\left(  \rho_{s}\right)  ,\rho_{w}\rangle
}{||t^{m}\left(  \rho_{s}\right)  ||\cdot||\rho_{w}||}$ The quadratic overhead
in the efficiency of the algorithm occurs if \ $f=\cos\gamma=1$ for
m=O($\sqrt{N}$ ). Indeed this is the case.

\begin{proposition} The radial and angular fidelities are respectively\\
$f=\frac
 {1}{4}\left[  1+ \cos^{m}\left(  2\psi\left(  x\right)  \right) 
\cos
 \phi\left(  \chi\right)  \right]  $ , and \ $\cos\gamma=\cos^{2}\frac
{\phi\left(  \chi\right)  }{2}$,\\ where \ $\frac{\phi\left(  \chi\right)  }%
{2}=m\psi\left(  \chi\right)  -m\theta\left(  x\right)  +\alpha$ , with
\ $\theta\left(  \chi\right)  =\pi+\chi+\sin^{-1}\left(  \frac{2\sqrt{N-1}}%
{N}\right)  $, and \ $\cos\alpha=\frac{1}{\sqrt{N}}$. For these figures of
merit there exist environments with parameters \ $\widehat{\chi}$
$>$%
0 such that , $\psi\left(  \widehat{\chi}\right)  =0,$ for which\ \ $f=\cos
\gamma=1,$\ \ when m=O($\sqrt{N}$ ). These values of \ $\widehat{\chi}$ are:
$\widehat{\chi}=$ $\chi_{n}$ $=$ $\pi\left(  4n^{2}-\frac{1}{4}\right)
^{\frac{1}{2}} $, for $\ n$ $\in$ $\mathbb{Z}_{+}$.

\end{proposition}

\section{Majorization and entropy increase in quantum searching
}

\begin{proposition} The map $\rho\rightarrow\rho^{^{\prime}}=t^{m}\left(
\rho\right)  $ ,$m$ $\in \mathbb{N}$ , majorizes [9,10] the vector \ $\lambda
_{\rho^{^{\prime}}}$ of eigenvalues of $\rho^{^{\prime}}$, by the vector
$\lambda_{\rho}$ of eigenvalues of $\ \rho$\ \ i.e. $\lambda_{\rho^{^{\prime}%
}}\prec\lambda_{\rho}$ \ and renders the dissipative search algorithm an
entropy (disorder) increasing one, in the sense that for the quantum
entropy\ $S\left(  k\right)  =-Tr\left(  k\log k\right)  $ is valid that
\ $S\left(  \rho^{^{\prime}}\right)  >S\left(  \rho\right)  $.

\end{proposition}

\section{Discussion
}

Grover's search algorithm together with its various extensions and
applications hold a prominent role in the flourishing field of quantum
algorithms, complexity and information. Here we have made a crucial test to
the algorithm. We question its efficiency when the omnipresent quantum noise
corrupts some of the ideal operation constituting the search map. If e. g
quantum phase damping is incorporated in the search operation, the ensuing
dissipative algorithm may exponentially fail in its speed of finding and the
accuracy of determining the target quantum state. Still there exist a
countable infinity of values of the damping parameter for which the noisy
algorithm is robust and performs its task quadratically faster than any
classical rival does. Finally, noisy searching creates in every step density
matrices that are majorized by the initial matrix, a thing that implies that
searching increases entropy if environment influences are taken into account.

Robustness of search algorithm for other types of quantum noise, as well as
entropy production, majorization and information aspects of quantum searching
are topics worth of future studying. Some of them are taken up in [11] where
detailed proofs of the statements of this paper will be given.

\end{document}